\documentclass[aps,prl,twocolumn,superscriptaddress,showpacs,preprintnumbers,amsmath,amssymb]{revtex4}
\usepackage{amssymb}
\usepackage[dvips]{graphicx}
\usepackage{color}

\begin{document}

\title{Excitonic Aharonov-Bohm Effect in Isotopically Pure $^{70}$Ge/Si Type-II Quantum Dots}

\author{Satoru~Miyamoto}
\affiliation{School of Fundamental Science and Technology, Keio University, 3-14-1 Hiyoshi, Kohoku-ku, Yokohama 223-8522, Japan}
\author{Oussama~Moutanabbir}
\affiliation{School of Fundamental Science and Technology, Keio University, 3-14-1 Hiyoshi, Kohoku-ku, Yokohama 223-8522, Japan}
\affiliation{Max Planck Institute of Microstructure Physics, Weinberg 2, D 06120 Halle (Saale), Germany}
\author{Toyofumi~Ishikawa}
\affiliation{School of Fundamental Science and Technology, Keio University, 3-14-1 Hiyoshi, Kohoku-ku, Yokohama 223-8522, Japan}
\author{Mikio~Eto}
\affiliation{School of Fundamental Science and Technology, Keio University, 3-14-1 Hiyoshi, Kohoku-ku, Yokohama 223-8522, Japan}
\author{Eugene~E.~Haller}
\affiliation{Lawrence Berkeley National Laboratory and UC Berkeley, 1 Cyclotron Rd., Berkeley, CA 94720, USA}
\author{Kentarou~Sawano}
\affiliation{Research Center for Silicon Nano-Science, Advanced Research Laboratories, Tokyo City University, 8-15-1 Todoroki, Setagaya-ku, Tokyo 158-0082, Japan}
\author{Yasuhiro~Shiraki}
\affiliation{Research Center for Silicon Nano-Science, Advanced Research Laboratories, Tokyo City University, 8-15-1 Todoroki, Setagaya-ku, Tokyo 158-0082, Japan}
\author{Kohei~M.~Itoh}
\email{kitoh@appi.keio.ac.jp}
\affiliation{School of Fundamental Science and Technology, Keio University, 3-14-1 Hiyoshi, Kohoku-ku, Yokohama 223-8522, Japan}
\date{\today}

\pacs{78.55.-m, 78.67.Hc, 71.35.Ji, 28.60.+s}

\begin{abstract}
We report on a magneto-photoluminescence study of isotopically pure $^{70}$Ge/Si self-assembled type-II quantum dots.  Oscillatory behaviors attributed to the Aharonov-Bohm effect are simultaneously observed for the emission energy and intensity of excitons subject to an increasing magnetic field.  When the magnetic flux penetrates through the ring-like trajectory of an electron moving around each quantum dot, the ground state of an exciton experiences a change in its angular momentum.  Our results provide the experimental evidence for the phase coherence of a localized electron wave function in group-IV Ge/Si self-assembled quantum structures.
\end{abstract}

\maketitle

The wave function of a charged particle circulating around a magnetic field region acquires a phase shift proportional to the magnetic flux threading the closed path.  This topological quantum effect, well known as the Aharonov-Bohm (AB) effect \cite{Aharonov59}, manifests as quantum interference of the particle wave function with the period of the universal flux quantum $\Phi_0 = hc/e$.  Tonomura \textit{et al}. provided the first conclusive experimental evidence for the existence of the AB effect by using magnetic fields shielded from an electron wave \cite{Tonomura86}.  Recent advances in lithography and growth techniques made it possible to observe the AB effect in the magneto-transport properties of semiconductor quantum rings within the experimentally available range of magnetic fields \cite{Gao94,Fuhrer01}.  In addition to this electronic AB effect, the possibility of the AB effect to occur for an exciton placed in a magnetic field was predicted assuming that the electron and hole move around with different ring-like trajectories \cite{Chaplik95,Kalameitsev98,Janssens01,Govorov02}.  Such an excitonic variety of the AB effect has been exclusively reported for compound semiconductor quantum structures such as InGaAs/GaAs patterned quantum rings \cite{Bayer03}, InAs/InP quantum tubes \cite{Tsumura07}, InP/GaAs and ZnTe/ZnSe self-assembled type-II quantum dots (QDs) \cite{Ribeiro04,Degani08,Kuskovsky07,Sellers08a,Sellers08b}.  In particular, the type-II QD enhances internal polarization of neutral exciton since the electron and hole are spatially separated.  The radial dipole moment accumulates the AB phase mainly during the coherent motion of either electron or hole that rotates around the QD in the presence of perpendicular magnetic fields.  This leads to a strong AB effect in the magneto-optical properties of such QD systems.\\
\indent In spite of being fundamentally and technologically highly relevant, such reports on Si-based quantum systems are conspicuously missing in the literature due to the relatively large effective mass and poor optical emission efficiency.  It is known that self-assembled Ge hut clusters in a Si matrix have small-size structures classified as type-II QDs \cite{Dashiell01,Denker03}.  In this Letter, we report a clear evidence of the AB effect on the bound excitons around this type of Ge/Si QDs.\\
\begin{figure}[b]
\includegraphics[width=7.0cm]{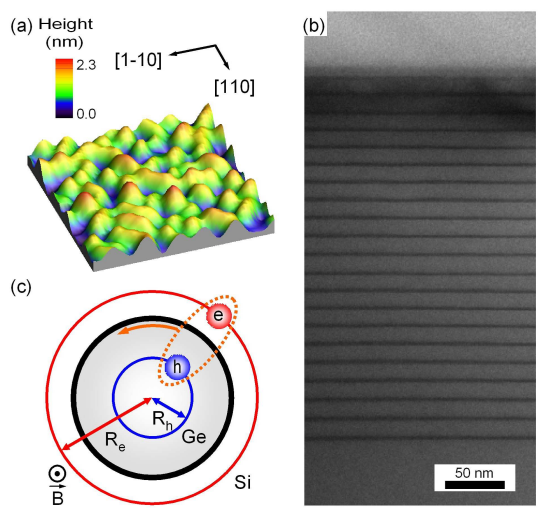}
\caption{\label{FIG1}(color).  (a) AFM image of the Ge/Si hut clusters formed by 6-ML Ge deposition at $540~^\circ$C (200 $\times$ 200~nm$^2$).  (b) Cross-sectional transmission electron microscope image of an as-grown superlattice comprising 20 periods of hut cluster layers separated by 15-nm Si layers.  (c) Schematic illustration of the spatial configuration of electron and hole in the Ge/Si type-II QD.}
\end{figure}
\indent A Ge/Si hut cluster superlattice was grown via the Stranski-Krastanov growth mode by solid-source molecular beam epitaxy (MBE).  A \textit{p}-type FZ-Si(001) substrate was chemically cleaned using the Ishizaka-Shiraki method \cite{Ishizaka86} before being introduced into the MBE chamber.  The protective oxide film was desorbed in an ultra high vacuum by annealing the substrate at $800~^\circ$C.  The thermal treatment was followed by 6-ML Ge deposition at the growth temperature of $540~^\circ$C.  An isotopically purified solid source of $^{70}$Ge possessing zero nuclear spin was used for the growth.  The low-temperature epitaxial growth resulted in the kinetic formation of defect-free hut clusters with both relatively small size distribution and high density in the order of 10$^{11}$~cm$^{-2}$.  Figure \ref{FIG1}(a) displays an image of typical morphology observed by \textit{ex-situ} atomic force microscope (AFM) in tapping mode.  The dimensions of the uncapped hut clusters were 31.8 $\pm$ 4.5~nm in length and width, and 1.3 $\pm$ 0.3~nm in height.  Immediately after the Ge deposition, the hut cluster layer was covered with a 15~nm-thick undoped Si spacer in order to reduce the vertical correlation which gives an additional factor for size dispersion \cite{Schmidt02}.  This stacking sequence was repeated to grow a 20-period QD superlattice that was free of visible dislocations [Fig. \ref{FIG1}(b)].  Residual defects unintentionally incorporated during the low-temperature growth \cite{Usami92} were predominantly annihilated by 1~s rapid thermal annealing (RTA) at $730~^\circ$C in an argon ambient \cite{Moutanabbir08}.  The average Ge composition within the QDs was estimated to be $x_{\textrm{Ge}} \sim 0.6$ by Raman scattering analysis.  The post-annealed sample was mounted in a strain-free manner and immersed in a superfluid liquid helium bath at the temperature of $T = 2$~K.  Low-temperature photoluminescence (PL) measurements were carried out by applying a magnetic field $B$ up to 5~T in the Faraday configuration.  The 514.5~nm line of an Ar$^+$ laser was employed as an excitation source.  The photogenerated holes are tightly trapped by the Ge-rich QDs while the electrons are weakly bound and extended into the Si regions [Fig. \ref{FIG1}(c)].  In addition to the Coulomb attraction, a built-in tensile strain in Si also creates an attractive potential for the electrons around QDs \cite{Dashiell01}.  The luminescence emission from the QDs was guided to a Bomem DA 8 Fourier transform interferometer having a spectral resolution of 250~$\mu$eV, and detected by a liquid nitrogen-cooled InSb photodetector.\\
\begin{figure}[t]
\includegraphics[width=8.0cm]{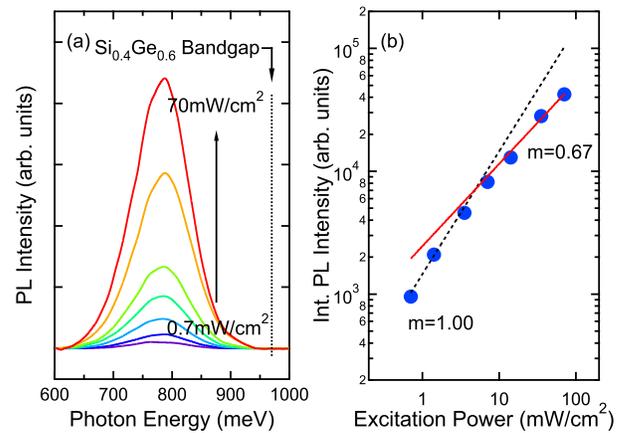}
\caption{\label{FIG2} (color).  (a) Zero-field PL spectra measured with the excitation power between 0.7 and 70~mW/cm$^2$.  The vertical dashed line indicates the bandgap position of the unstrained Si$_{0.4}$Ge$_{0.6}$ bulk.  (b) Power dependence $I_{\textrm{PL}} \propto P_{\textrm{exc}}^m$ of the intensity obtained by integration of the PL spectra.  The power exponent asymptotically changes from $m = 1.00$ (dotted line) to $m = 0.67$ (solid line).}
\end{figure}
\indent Figure \ref{FIG2}(a) shows the zero-field PL spectra recorded with varying excitation powers.  The spectra extend to energies below the band gap of the unstrained Si$_{0.4}$Ge$_{0.6}$ bulk \cite{Weber89}, supporting the observation that the grown hut clusters have the type-II band alignment.  We observe no clear PL features related to dislocations \cite{Usami95} and impurities which can lead to a modification in the excitonic AB oscillations \cite{DiasdaSilva04,Degani08}.  Remarkably, the integrated PL intensity displays a sublinear power dependency on the excitation power above 7~mW/cm$^2$ [Fig. \ref{FIG2}(b)].  The asymptotical power exponent of $m \sim 0.67$ has been predicted when Auger recombination process exists as a competitive nonradiative channel \cite{Apetz95}.  In such cases, the PL energy is expected to show a blue shift due to the many-body effects \cite{Kamenev06}.  The present experimental results show the same phenomena as observed for PL properties of the type-II Ge/Si hut clusters with high excitation powers \cite{Dashiell01}.  Meanwhile, a linearly dependent regime can be observed at weaker excitation power, which is difficult to be achieved in the single-layer structure aforementioned.  This is possibly attributed to the experimental fact that the Auger recombination process is sufficiently suppressed owing to the use of the multi-layer structure \cite{Tayagaki09}.  The excitation power density of 3.5~mW/cm$^2$ used for the observation of the excitonic AB oscillations populates less than one exciton per QD.  During a relatively long recombination time caused by a large wave function separation \cite{Kamenev06}, most of single excitons relax to the ground states.\\
\indent The tensile strain around the QD splits the six-fold $\Delta$ valley degeneracy of the Si conduction band into twofold and fourfold valleys.  The low-lying twofold valleys in the growth direction are mainly occupied by electrons.  Since the intense luminescence is here observed despite of the indirect nature of the band gap [Fig. \ref{FIG2}(a)], the momentum of the electron is likely to be no longer a good quantum number in the growth direction.  The coupling between the two valleys in the momentum space induces nondegenerate states around the $\Gamma$ point, thereby allowing phononless radiative transition \cite{Dashiell01}.  Recently, the valley splitting of the order of 1~meV was observed by lateral confinement of the electron wave functions in Si/SiGe quantum well \cite{Goswami07}.  Although the valley splitting in our Ge/Si quantum dots may be larger than the case for the quantum well due to the strong confinement and strain in the growth direction, such an energy splitting is covered by the inhomogeneous broadening in the present experiment.\\
\indent Figure \ref{FIG3} shows the PL spectra of Ge/Si hut cluster superlattice recorded at two different magnetic fields, $B = 0.8$~T and $B = 1.8$~T.  The PL intensity is reduced with increasing the magnetic fields.  The spectra are fitted well with single Gaussian curves, from which the peak positions are determined.  The application of the magnetic fields induces an overall peak shift in the order of a few meV.  Figures \ref{FIG4}(b)-(d) show the detailed evolution of the PL intensity and the peak energy as a function of the magnetic field.  Oscillatory behaviors are observed for both energy and intensity providing a clear signature of the excitonic AB effect.\\
\begin{figure}[t]
\includegraphics[width=6.0cm]{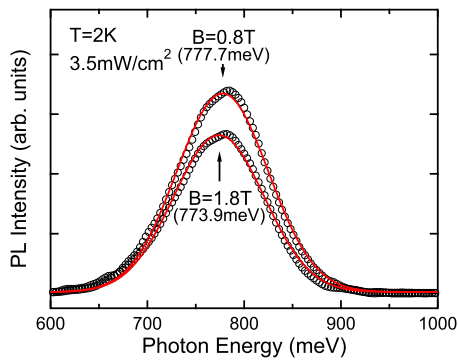}
\caption{\label{FIG3}(color).  PL spectra of the Ge/Si hut cluster superlattice recorded at 2~K for the two magnetic fields, $B = 0.8$~T and $B = 1.8$~T.  In order to extract the peak positions, the experimental data are fitted with single Gaussian curves (solid lines).  The arrows indicate the fitted peak energy.}
\end{figure}
\begin{figure}[t]
\includegraphics[width=6.75cm]{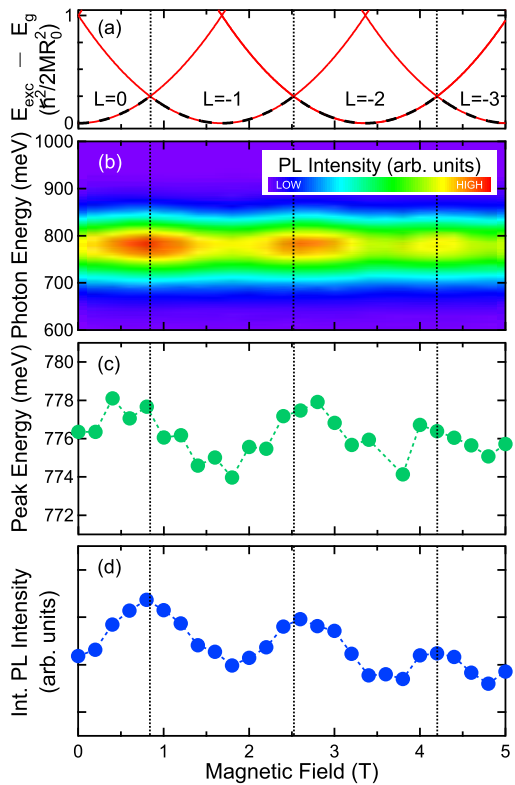}
\caption{\label{FIG4}(color).  (a) Magnetic-field dependence of the quadratic contribution in Eq. (\ref{Eq1}) representing the excitonic energy for an angular momentum $L$.  Although each $L$ state consists actually of two nondegenerate states considering the valley degree of freedom, only the lower energy state of those states is shown for simplicity.  The dashed line marks the energetic transition for ground-state excitons.  (b) Contour plot of the PL spectra as a function of the magnetic fields.  Detailed evolutions of (c) the peak energy and (d) the integrated PL intensity.  The vertical dotted lines indicate the magnetic fields at which the angular momentum transition takes place.  The magnetic-field separation between the neighboring dotted lines corresponds to the flux quantum $\Phi_0$.}
\end{figure}
\indent The origin of the oscillation of the peak energy can be explained as follows.  The energy of the single exciton with a definite angular momentum (projection), $L$, is described by \cite{Govorov02}:
\begin{equation}
E_{\textrm{exc}}(B) = E_g + \frac{\hbar^2}{2MR_0^2}\left(L + \frac{\Delta\Phi}{\Phi_0}\right)^2\label{Eq1}
\end{equation}
where $E_g$ is the energy gap including the exciton binding energy.  $M = (m_eR_e^2 + m_hR_h^2)/R_0^2$ and $R_0 = (R_e + R_h)/2$ are defined for describing the rotation Hamiltonian of the whole exciton along the ring-like trajectory (see Fig. \ref{FIG1}(c)), where $m_{e(h)}$ and $R_{e(h)}$ are the effective mass and orbital radius of the electron (hole), respectively.  Assuming that the hole is localized strongly within the QD, i.e. $R_h \approx 0$, the magnetic flux threading the region between the electron and hole orbital rings is given by $\Delta\Phi = \pi R_e^2B$.  As illustrated in Fig. \ref{FIG4}(a), the second term in Eq. (\ref{Eq1}) shows the quadratic dependence of the excitonic energy on the magnetic field for different values of $L$.  At zero magnetic field, the ground state of the exciton takes a zero angular momentum $L = 0$.  With increasing the magnetic fields, the angular momentum makes a series of ground-state transition to $L = -1$, $-2$, and $-3$ for the half integer flux quanta.  Then the energy of the ground-state exciton oscillates with the magnetic-field period of $\Phi_0/\pi R_e^2$ along the dashed line in Fig. \ref{FIG4}(a).  This qualitative account is in good agreement with the oscillation of the peak energy observed in Fig. \ref{FIG4}(c).  We believe that a certain level of damping is due to inhomogeneity associated with the QD ensembles similarly to the previous observation in InP/GaAs system \cite{Ribeiro04}.  It should be noted that the absolute value of the angular momentum $|L|$ denotes the number of threading flux quanta.  In Fig. \ref{FIG4}(c), the oscillation is sustained up to magnetic fields corresponding to three flux quanta.  From the period of AB oscillations, the average radius of coherent electron motion is estimated to be $R_e \sim 28$~nm.  This orbital diameter, $2R_e$, is slightly larger than the structural size measured by AFM although unavoidable intermixing during the Si overgrowth and RTA may modify the Si-Ge interface.\\
\indent Figure \ref{FIG4}(d) shows an oscillatory behavior of the PL intensity that is in phase with the oscillation of the peak energy.  However, the simplified picture of the electron travelling with the constant orbital radius of $R_e$ cannot explain the oscillation of the PL intensity.  As a matter of fact, increasing the magnetic field tends to push the electron wave function closer to the QD boundary \cite{Janssens01,Silva06}.  In case of a decrease in the diameter, it can enlarge the overlap of the electron and hole wave functions, thereby resulting in the monotonic enhancement of the PL intensity.  On the other hand, the angular momentum transition from $L$ to $L-1$ should accompany the expansion of the electron trajectory \cite{Janssens01,Silva06} and suppress the PL intensity abruptly.  Hence the interplay between localization and delocalization of electrons periodically switches the PL intensity.  However, the intensity oscillation shown in Fig. \ref{FIG4}(d) does not necessarily represent a saw-tooth shape but a sinusoidal one, which can be attributed to the fact that the excitonic populations are thermally distributed in the ground states as well as the excited states at finite temperature [Fig. \ref{FIG4}(a)].  In principle, the selection rule that the optically bright transitions are allowed only for the excitons with $L = 0$ is valid in nanostructures with a perfect rotational symmetry.  Due to the asymmetric shape of typical hut clusters elongated in [100] directions (see Fig. \ref{FIG1}(a)), the partial relaxation of the selection rule activates the dark excitons with $L \ne 0$ \cite{Degani08,DiasdaSilva04}.  The amplitude of the PL intensity oscillation thus decreases as a general trend but persists even after each angular momentum transition.\\
\indent In conclusion, we have presented the AB signature of the ground-state excitonic energy through magneto-PL measurements on Ge/Si self-assembled type-II QDs.  The localized electrons in the tensile-strained Si surrounding region display sufficient coherence to exhibit an oscillatory behavior of the PL intensity that is in phase with that of the excitonic energy.  Such synchronized oscillations cannot be understood in the framework of a simplified theory assuming the constant diameter of the electron motion around the QD.  Although further investigations are necessary to comprehend the oscillatory behaviors, these oscillations suggest the possibility to tailor the optical properties of Si-based nanostructures by a controlled application of magnetic fields.\\
\indent The authors acknowledge A. Sagara, H. Oshikawa, and K. Yoshizawa for their technical supports.  This work was supported in part by Grant-in-Aid for Scientific Research by MEXT (Specially Promoted Research No. 18001002), in part by Special Coordination Funds for Promoting Science and Technology for INQIE, and in part by Grant-in-Aid for the Global Center of Excellence for High-Level Global Cooperation for Leading-Edge Platform on Access Spaces from MEXT.

\end{document}